\RequirePackage{silence}
\documentclass[reprint,amsmath,amssymb,aps,pra,superscriptaddress,nofootinbib, floatfix, citeautoscript,
]{revtex4-2}
\usepackage[utf8]{inputenc}
\usepackage[english]{babel}
\usepackage{hyperref}
\usepackage{mathtools}
\usepackage{bm}
\usepackage{hyperref}
\usepackage{float}
\usepackage{physics}
\usepackage{tabularx}
\usepackage[page]{appendix}
\usepackage{nicefrac}
\usepackage{wrapfig}
\usepackage[per-mode=power, 
            exponent-product = \times, 
            inter-unit-product = \ensuremath{{\cdot}}, 
            tight-spacing = true,
            number-unit-product=~,
            parse-numbers=false,]{siunitx}
\AtBeginDocument{\RenewCommandCopy\qty\SI}
\ExplSyntaxOn
\msg_redirect_name:nnn { siunitx } { physics-pkg } { none }
\ExplSyntaxOff
\usepackage{multibib}
\usepackage[dvipsnames]{xcolor}
\usepackage{booktabs}
\usepackage{float}
\usepackage{graphicx}
\usepackage[capitalise]{cleveref}
\usepackage[version=4]{mhchem}
\usepackage{tabularx}
\usepackage{ulem}
\usepackage{enumitem}
\usepackage{xcolor}
\usepackage[
    protrusion=true,
    activate={true,nocompatibility},
    final,
    spacing=nonfrench,
    tracking=true,
    kerning=true,
    spacing=true,
    factor=1000]{microtype}
\hypersetup{
 colorlinks=true, 
 linkcolor=blue, 
 citecolor=blue,        % color of links to bibliography
 filecolor=blue,      % color of file links
 urlcolor=blue}
\usepackage[raggedright]{titlesec}
\usepackage{caption}
\titleformat*{\section}{\bfseries\large}
\titleformat*{\subsection}{\bfseries}
\titlespacing*{\section}{0em}{0.75em}{0pt}
\renewcommand{\textcite}[1]{\citenum{#1}}
\newcolumntype{Y}{>{\centering\arraybackslash}X}
\Crefformat{appendix}{Supplementary Information #2#1#3}

\begin{document}
%               _    _                  
%   __ _  _  _ | |_ | |_   ___  _ _  ___
%  / _` || || ||  _|| ' \ / _ \| '_|(_-<
%  \__,_| \_,_| \__||_||_|\___/|_|  /__/
                                      
\author{Gr\'egory~Moille}%
\email{gmoille@umd.edu}
\affiliation{Joint Quantum Institute, NIST/University of Maryland, College Park, USA}
\affiliation{Microsystems and Nanotechnology Division, National Institute of Standards and Technology, Gaithersburg, USA}
\author{Pradyoth~Shandilya}
\affiliation{University of Maryland, Baltimore County, Baltimore, MD, USA}
\author{Miro~Erkintalo}
\affiliation{Department of Physics, University of Auckland, Auckland 1010, New Zealand}
\affiliation{The Dodd-Walls Centre for Photonic and Quantum Technologies, New Zealand}
\author{Curtis~R.~Menyuk}
\affiliation{University of Maryland, Baltimore County, Baltimore, MD, USA}
\author{Kartik~Srinivasan}
\affiliation{Joint Quantum Institute, NIST/University of Maryland, College Park, USA}
\affiliation{Microsystems and Nanotechnology Division, National Institute of Standards and Technology, Gaithersburg, USA}

\date{\today}

%   _    _  _    _      
%  | |_ (_)| |_ | | ___ 
%  |  _|| ||  _|| |/ -_)
%   \__||_| \__||_|\___|
% \title{Metrological Binding of Integrated Cavity Solitons via Kerr-Induced Synchronization }

\newcommand{\mytitle}{On-Chip Parametric Synchronization of a Dissipative Kerr Soliton Microcomb}
\title{\mytitle}
%         _         _                   _   
%   __ _ | |__  ___| |_  _ _  __ _  __ | |_ 
%  / _` || '_ \(_-<|  _|| '_|/ _` |/ _||  _|
%  \__,_||_.__//__/ \__||_|  \__,_|\__| \__|
\begin{abstract}
    %112 words 783 charac (limt 600)
    Synchronization of oscillators is ubiquitous in nature. Often, the synchronized oscillators couple directly, yet in some cases synchronization can arise from their parametric interactions. 
    Here, we theoretically predict and experimentally demonstrate the parametric synchronization of a dissipative Kerr soliton frequency comb. We specifically show that the parametric interaction between the soliton and two auxiliary lasers permits the entrainment of the frequency comb repetition rate. Besides representing the first prediction and demonstration of parametric synchronization of soliton frequency combs, our scheme offers significant flexibility for all-optical metrological-scale stabilization of the comb.
\end{abstract}

\maketitle
% figure word equivalent = 978
% equation word equivalent = 128
% to add -> 1106

% MAX word for PRL: 3,750
%08/01/2024: 
%   words = 2499
% ***total = 3,605*** YEAH!!!

%   ___       _             
%  |_ _| _ _ | |_  _ _  ___ 
%   | | | ' \|  _|| '_|/ _ \
%  |___||_||_|\__||_|  \___/
\textit{Introduction---}%
Synchronization is ubiquitous in nature, from coupled pendulums~\cite{Strogatz2004sync} to fireflies~\cite{BuckScience1968}, neurons~\cite{FellNatRevNeurosci2011}, and quantum systems~\cite{JainPhysicsReports1984,MilitelloPhys.Rev.A2017}. Despite their drastic differences, these systems' synchronization dynamics typically follow common universal patterns and are, to first-order, governed by the same mathematical equations. In optics, the same is true for dissipative Kerr solitons (DKSs), which are cornerstones for the creation of on-chip frequency combs~\cite{KippenbergScience2018}. Synchronization between DKSs has been demonstrated, for instance, between counter-propagative solitons~\cite{YangNaturePhoton2017}, or solitons existing in remote resonators~\cite{JangNaturePhoton2018}. Recently, It has also been shown that a DKS can synchronize to an external continuous-wave reference optical field~\cite{MoilleNature2023, WildiAPLPhotonics2023}, following the same Adler model as coupled oscillators~\cite{MoilleNature2023,CoulletAmericanJournalofPhysics2005}.
  In this Kerr-induced synchronization (KIS) regime, the phase locking of the DKS results in the capture of one comb tooth by the reference field~\cite{MoilleNature2023,Wildi2024}. Since the main pump creating the DKS is also a comb tooth, KIS provides a passive dual-pinning of the DKS frequency comb, enabling low-noise operation of the microcomb below the fundamental limit imposed by the resonator thermorefractive noise~\cite{Moille2024_arXivTRN}, which is critical for metrology applications such as timekeeping~\cite{CundiffRev.Mod.Phys.2003}, time-transfer~\cite{CaldwellNature2023a}, ranging~\cite{ZhangAppliedPhysicsLetters2016}, or spectroscopy~\cite{BjorkScience2016,HolzwarthPhys.Rev.Lett.2000}. %
  Although KIS can occur at any comb tooth~\cite{WildiAPLPhotonics2023}, efficient synchronization requires that the reference laser is both close to a comb line and on-resonance, which is challenging to achieve simultaneously due to dispersion, particularly for large frequency separations between the main and reference pumps, which is desirable for optical frequency division (OFD) and clockworks~\cite{MoilleNature2023}.\\
\indent In this work, we leverage a Kerr parametric interaction driven by two reference lasers to obtain a new DKS synchronization regime that bypasses the above limitation. Related parametric processes have recently attracted significant attention, e.g., for all-optical random number generation~\cite{OkawachiOpt.Lett.OL2016} and optical spin-glasses~\cite{OkawachiNatCommun2020}, or for obtaining a new type of parametrically-driven dissipative soliton~\cite{MoilleNat.Photon.2024}. However, this type of parametric interaction has yet to be explored in the context of DKS synchronization. We show for the first time that the interaction between two on-resonance auxiliary lasers, outside of the DKS comb frequency grid, along with the DKS comb itself, can yield a parametric driving force for the soliton that mediates synchronization. We theoretically unveil the conditions for efficiently obtaining this ``parametric-KIS'', finding that the resonator must exhibit at least third-order dispersion to support a zero crossing of the integrated dispersion.
Experimentally, we demonstrate this effect using an octave-spanning comb in a \ce{Si3N4} microring resonator. Similar to standard KIS~\cite{MoilleNature2023}, parametric-KIS stabilizes the microcomb, such that its repetition rate becomes dependent on the three lasers at play.

\begin{figure}[t]
    \begin{center}
        \includegraphics[width = \columnwidth]{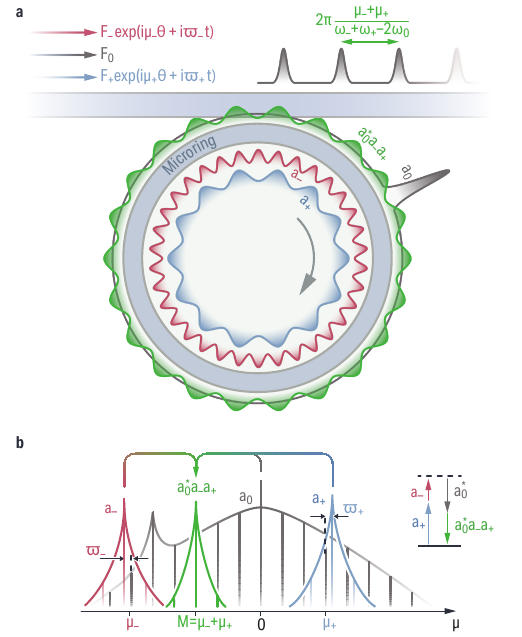}
        % aspect ratio (width / height) 0.8190360815
        % word equivalent [150 / (aspect ratio)] + 20
        % = 203
        \caption{\label{fig:1}
        \textbf{a} Schematic of the parametric-KIS system. A main pump $F_0$ generates a DKS $a_0$. Two additional drives $F_\pm$ at mode indices $\mu_\pm$ and relative frequency offset $\varpi_\pm$ from their nearest DKS comb tooth are injected into the same resonator to create the two other colors $a_-$ and $a_+$. The parametric interaction between the three colors $a_0$, $a_-$, and $a_+$ creates a parametric drive for $a_0$, which when in phase with the DKS can synchronize it. Thus, it disciplines the pulse train repetition rate. %
        \textbf{b} Spectral representation of the same system as in (a) where each comb component is frequency offset by $\varpi_\pm$ respectively, highlighting how  the parametric drive at $M = \mu_- + \mu_+$ provides the optical frequency division factor.
        }
    \end{center}
\end{figure}

%   _____  _                         
%  |_   _|| |_   ___  ___  _ _  _  _ 
%    | |  | ' \ / -_)/ _ \| '_|| || |
%    |_|  |_||_|\___|\___/|_|   \_, |
%                               |__/ 
\vspace{1em}
\textit{Results---}
First, we present the theoretical framework of the novel parametric-KIS scheme. The system consists of a microring resonator that is triply pumped [\cref{fig:1}a]. The intracavity field $a(\theta, t)$ can be modeled using using a modified Lugiato-Lefever equation (mLLE)~\cite{TaheriEur.Phys.J.D2017}:
\begin{equation}
    \begin{split}
    \label{eq:mLLE} 
    \frac{\partial a(\theta,t)}{\partial t} &= \left(-\frac{\kappa}{2} + i\Delta \omega_\mathrm{0}\right)a 
        + i\sum_\mu D_\mathrm{int}(\mu)A(\mu,t)\mathrm{e}^{i\mu\theta} \\
        &- i\gamma |a|^2 a + iF_0 \\
        & + iF_-\mathrm{e}^{i\varpi_-t + i\mu_\mathrm{-}\theta}
        + iF_+\mathrm{e}^{i\varpi_+t + i\mu_\mathrm{+}\theta}
    \end{split}
    \raisetag{3em}
\end{equation}
\noindent where $\theta$ is the azimuthal coordinate that rotates with the DKS angular group velocity, $t$ is time, $\mu$ is the mode difference with respect to the mode of the primary pump, and $A(\mu,t)$ is the Fourier transform of $a(\theta,t)$.  The parameters $\kappa$, $\Delta\omega_0 = \omega_\mathrm{res}(0) - \omega_0 $, and $\gamma$ denote the total loss rate, the offset between the primary pump $\omega_0$ and the primary pump resonance $\omega_\mathrm{res}(0)$, and the Kerr coefficient, respectively. The parameter $F_0$ is related to the primary pump power $P_0= F_0^2/\kappa_\mathrm{ext}$, where $\kappa_\mathrm{ext}$ is the coupling loss rate.  We similarly define $F_\pm$ via the relations $P_\pm =F_\pm^2/\kappa_\mathrm{ext}$. The modified integrated dispersion, which we define with the DKS repetition rate outside of synchronization $\omega_\mathrm{rep}^{(0)}$ instead of the angular free spectral range around the pump resonance $D_1$, is $D_\mathrm{int}(\mu) = \omega_\mathrm{res}(\mu) - \left( \omega_\mathrm{res}(0) +\mu \omega_\mathrm{rep}^{(0)} \right)= \left( D_1-\omega_\mathrm{rep}^{(0)} \right)\mu + \sum_{k>1} D_k\mu^k/k!$, where $\omega_\mathrm{res}(\mu)$ is the frequency of resonance at mode $\mu$, and $D_k$ the higher order dispersion terms. The two auxiliary pumps are at frequencies $\omega_\pm$ and are located at modes $\mu_\pm$ with respect to the primary pump, such that $\mu_-<0$ and $\mu_+>0$. These pumps are     offset from their nearest comb tooth by $\varpi_\pm =  \omega_\pm - \omega_0 - \mu\omega_\mathrm{rep}^{(0)}$.
  
In the regime of interest to us here, the integrated dispersion is sufficiently small for a three-component multi-color soliton (McS) to form~\cite{MoilleNat.Commun.2021,3wave}.  
The McS consists of the DKS and two azimuthally localized structures with carrier frequencies $\omega_\pm$ that are locked to each other in group (but not phase) velocity, leading to different accumulated phase shifts at each round trip. In the frequency domain, the McS is associated with three interleaved frequency combs that share the same repetition rate $\omega_\mathrm{rep}$ but are offset from one another, with $\varpi_\pm$ representing the offset of the combs around the auxiliary pump frequencies from the DKS comb [\cref{fig:1}b].
%The McS is characterized within the microresonator, where there is an azimuthally locked structure composed of the DKS and the auxiliary pump sharing the same group velocity, but with different accumulated phase shifts at each round trip. As a result, it produces three distinct comb components at the waveguide output, sharing the same repetition rate $\omega_\mathrm{rep}$ but interleaved with an offset from one another of $\varpi_\pm$ [\cref{fig:1}b]. %a single group velocity or repetition rate frequency $\omega_\mathrm{rep}$, but by different phase velocities for each of the three comb components [\cref{fig:1}b]. As a result, it produces interleaved frequency combs with different offset frequencies $\varpi$.  
Standard KIS is achieved by tuning $\varpi_+$ or $\varpi_-$ to be small, such that the corresponding pump captures a comb tooth, at which point the colors generated by the two pumps become indistinguishable~\cite{MoilleNature2023}. In stark contrast, in the parametric-KIS regime explored in this work, the parameters $\varpi_\pm$ are generally large such that standard KIS does not occur. In this case, and similar to ref.~\cite{3wave}, the total intracavity field can be expanded as a superposition of the three colors viz.
%When either $\varpi_+$ or $\varpi_-$ becomes sufficiently small, it is possible for the corresponding pump to capture a comb tooth, at which point the colors generated by the two pumps become indistinguishable~\cite{MoilleNature2023}.  However, we will not be treating that case here.  Hence, similar to ref.~\cite{3wave}, we may write the total field as:
\begin{equation}
    \label{eq:colorsplit}
    \begin{split}
        a(\theta,t)=& a_\mathrm{0}(\theta,t) 
        + a_\mathrm{-}(\theta,t)\mathrm{e}^{\left( i\varpi_- t + \mu_-\theta \right)}\\
        &+ a_\mathrm{+}(\theta,t)\mathrm{e}^{i\left( \varpi_+ t + \mu_+\theta \right)}
    \end{split}
\end{equation}
\indent After some algebra detailed in Supplementary Information S.1, \cref{eq:mLLE,eq:colorsplit} lead to the DKS equation: 
\begin{equation}
    \begin{split}
        \label{eq:a0} 
        \frac{\partial a_0(\theta,t)}{\partial t} &= \left(-\frac{\kappa}{2} +i\Delta \omega_\mathrm{0}\right)a_0 
            + i\sum_\mu D_\mathrm{int}(\mu)A_0(\mu, t)\mathrm{e}^{i\mu\theta} \\
            &-i\gamma \left(2|a_-|^2 + |a_0|^2 + 2|a_+|^2\right) a_0 \\
            & -i2\gamma a_0^*a_+a_- \mathrm{e}^{i\left(Wt + M\theta \right)} + iF_0
        \end{split}
        \raisetag{3em}
\end{equation}
where $W = \varpi_{-} + \varpi_{+}$ is the frequency offset of the idler wave that is generated via the parametric interaction from its closest DKS comb line at mode $M = \mu_{-} + \mu_{+}$.  Equation~(\ref{eq:a0}) is similar to the master equation of the $\chi^{(3)}$-mediated parametric soliton~\cite{MoilleNat.Photon.2024}, except with an additional direct driving force $F_0$; \cref{eq:a0} is also similar to the equation used to study standard-KIS of DKSs~\cite{MoilleNature2023}, but now with a parametric synchronization term $2\gamma a_0^*a_+a_-$ from the four-wave mixing between the reference fields $a_\pm$ and the soliton $a_0$, which concomitantly generates an idler field at $\omega_+ + \omega_- - \omega_0$. Hence, we may anticipate that the DKS in the triply-driven scheme shown in \cref{fig:1}(a) can experience synchronization, provided that the parametric driving term is sufficiently close in phase with the DKS. Indeed, a detailed analysis shows that, similar to any other synchronization mechanism for coupled oscillators, parametric-KIS obeys an Adler equation (see~Supplementary Information S.2), where $W$ is compensated by a temporal phase-slip of the DKS to achieve phase locking. %:
The analysis (detailed in the Supplementary Information S.1) leading to \cref{eq:a0} also shows that, when parametrically-synchronized, the comb exhibits the OFD factor $M=\mu_{-} + \mu_{+}$ which, contrary to direct-KIS, arises from the nonlinear interaction between the three colors. Therefore, parametric-KIS enables for a triple-pinning of the repetition rate from degenerated four-wave mixing, in contrast with the direct-KIS dual-pinning:
\begin{equation}
    \label{eq:triple_pinning}
    \omega_\mathrm{rep}^\mathrm{(pkis)} = \frac{\omega_{-} + \omega_{+} - 2\omega_0}{M}    \quad.
\end{equation}

Although off-resonance operation of the auxiliary pumps (as in direct-KIS~\cite{WildiAPLPhotonics2023}) is possible, the efficiency of parametric-KIS is optimized when the pumps are on resonance to maximize their respective intracavity powers, while allowing for $|W|$ to be minimized. Such condition exists if $D_\mathrm{int}(\mu_-) \approx  - D_\mathrm{int}(\mu_+)$.  Thus, to achieve on-resonance parametric-KIS operation, the resonator should have at least one zero crossing in its integrated dispersion $D_\mathrm{int}(\mu)$, and therefore should exhibit at least a non-negligible $D_3$ term.

%   ___  _        
%  / __|(_) _ __  
%  \__ \| || '  \ 
%  |___/|_||_|_|_|

\begin{figure}[t]
    \includegraphics[width=\columnwidth]{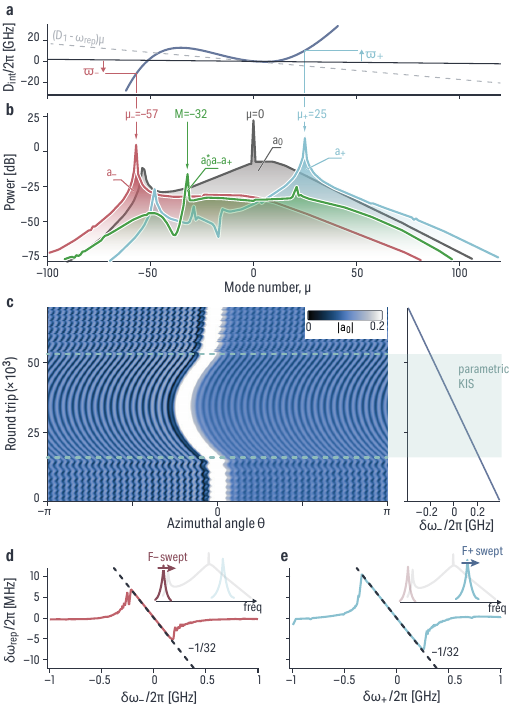}
    % aspect ratio (width / height) 0.7004555809
        % word equivalent [150 / (aspect ratio)] + 20
        % = 234
    \caption{\label{fig:2}
    \textbf{a} Modifed integrated dispersion $D_\mathrm{int}(\mu)$ of the resonator used in the simulation. %
    \textbf{b} Simulated comb spectrum for the three colors at play: the DKS' $a_0$ (grey) and both auxiliaries' $a_\pm$ with $\mu_-=-57$ (red) and $\mu_+=25$ (blue) %that have been chosen such that at zero auxiliary pumps' detuning, $W$ is minimized, which then can be zeroed through small detuning of one of the auxiliary pumps. %
    The resulting parametric driving field for the DKS' color $a_0$ is then at $M = \mu_- + \mu_+ = -32$ (green). %
    \textbf{c} Azimuthal profile (left, colorscale) of the DKS color with respect to the negative auxiliary pump detuning $\delta\omega_-$ (right). Outside of KIS, the parametric drive has an offset leading to a phase slip in time relative to the DKS, resulting in a CEO offset in the frequency domain. Once synchronized, their phase velocities lock and the variation of $W$ entrains the DKS's and hence disciplines its repetition rate, as apparent through its azimuthal drift.
    \textbf{d-e} Repetition rate variation $\delta \omega_\mathrm{rep}$ of the DKS with respect to the detuning of the negative (d) and positive (e) auxiliary pump detuning $\delta\omega_-$ and $\delta\omega_+$ respectively. Once the parametric KIS is reached, $\delta\omega_\mathrm{rep}$ varies with the OFD $M = \mu_- + \mu_-$, accordingly with~\cref{eq:triple_pinning}
    }
\end{figure}

\vspace{1em}
We demonstrate parametric-KIS numerically by solving~\cref{eq:a0} along with the equation describing $a_-$ and $a_+$ (see Supplementary Information S.1), using an integrated dispersion described by a cubic function with \qty{D_2/2\pi = 21}{\MHz} and \qty{D_3/2\pi = 1.25}{\MHz}, and assuming \qty{D_1/2\pi = 983.346}{\GHz}. The DKS with \qty{\omega_\mathrm{rep}/2\pi = 983.515}{GHz}, different from $D_1$ because of dispersive wave (DW) recoil~\cite{CherenkovPhys.Rev.A2017}, is generated by a main pump with power \qty{P_0=150}{mW} and detuning \qty{\Delta\omega_0/2\pi = -3.2}{GHz} in a system where \qty{\kappa_\mathrm{ext} = \kappa/2 = 2\pi\times200}{MHz} for critical coupling condition. The modified integrated dispersion presents a zero crossing at $\mu = -53$ [\cref{fig:2}a], where $a_0$ exhibits the creation of a dispersive wave. %
We choose the auxiliary pumps, with power \qty{P_-=1}{mW} and \qty{P_+=3}{mW} respectively, at $\mu_- = -57$ and $\mu_+=25$ for the components $a_-$ and $a_+$ to exhibit, while on resonance, relative offsets that are almost equal with opposite signs \qty{-\varpi_-/2\pi \approx \varpi_+/2\pi \approx 10}{\GHz}. 
By fine tuning the frequency detunings $\delta\omega_\pm$ of the auxiliary pumps, the frequency offset $W$ between the parametrically generated idler field and a DKS comb tooth can be pushed to lie
%, which from their $\delta \omega_\pm$ tuning can lead to $W$ 
within the parametric-KIS bandwidth (\textit{i.e.} $\lesssim1$~GHz).
%  The obtained DKS comb spectrum for $a_0$ has a dispersive wave (DW) at $D_\mathrm{int}(\mu_{DW}) = -53$ [\cref{fig:2}b], offset from the $D_\mathrm{int}$ zero crossing because of the dispersion asymmetry and soliton recoil~\cite{CherenkovPhys.Rev.A2017}.
  Additionally, the colors $a_-$ and $a_+$, which are not phase synchronized with $a_0$, experience the creation of additional azimuthal tones, regardless of their dispersion regime, thanks to the group-velocity binding of all the colors through cross-phase modulation (XPM)~\cite{WangOptica2017, QureshiCommunPhys2022, MoillearXiv2024}. The parametric driving force that synchronizes the DKS and which results from the three colors can be numerically extracted, exhibiting a clear tone at $M = \mu_{-} + \mu_{+} = -32$ at the $a_0$ color i.e. $|W|\ll|\varpi_\pm|$, as expected by the theory. %
The azimuthal profile of $a_0$ with respect to time allows us to understand the synchronization mechanism [\cref{fig:2}c]. Outside of synchronization, the parametric term exhibits a phase slip $W t \pmod{2\pi}$ from the DKS, resulting in a non-stationary interference pattern, equivalent to a CEO offset in the frequency domain. Once the parametric KIS is reached and synchronization is achieved, the $a_0^*a_-a_+$ driving term becomes in phase with the DKS, hence the absence of modulation of the azimuthal profile in time.\\
\indent While synchronized, a change in frequency of either auxiliary pump leads to a phase change of the synchronized DKS, causing a drift in its position in time and thus its repetition rate. However, none of the auxiliary pumps directly capture any comb teeth, which is in stark contrast with direct-KIS~\cite{MoilleNature2023} or other color-KIS~\cite{MoillearXiv2024} schemes. We can extract the repetition rate change $\delta\omega_\mathrm{rep}$ of $a_0$, which through XPM is the same as $a_\pm$, and study its entrainment with the auxiliary pumps' frequency change $\delta\omega_\pm$. As expected from \cref{eq:triple_pinning} in the parametric-KIS regime,  we obtain $\delta \omega_\mathrm{rep}/\delta\omega_- = \delta\omega_\mathrm{rep}/\delta \omega_+ = 1/M$ [\cref{fig:2}d-e]. We also confirm such results with different $\mu_\pm$ combinations (see Supplementary Information S.6), where unlike direct KIS, which allows efficient resonant operation only at the DW, parametric-KIS allows it at any $\mu_\pm$ pairs where $D_\mathrm{int}(\mu_-) = -D_\mathrm{int}(\mu_+)$. In this context, parametric-KIS enables more efficient synchronization at the off-resonance comb tooth $M$ than direct-KIS with comparable reference powers (see Supplementary Information S.8). This occurs due to the the near-resonant input fields driving an efficient non-degenerate four-wave mixing process that produces large intracavity power at an (off-resonant) idler frequency to capture that $M^\text{th}$ comb tooth.

%   ___                       _                   _   
%  | __|__ __ _ __  ___  _ _ (_) _ __   ___  _ _ | |_ 
%  | _| \ \ /| '_ \/ -_)| '_|| || '  \ / -_)| ' \|  _|
%  |___|/_\_\| .__/\___||_|  |_||_|_|_|\___||_||_|\__|
%            |_|                                      
\begin{figure*}
    \begin{center}
        \includegraphics[width = \textwidth]{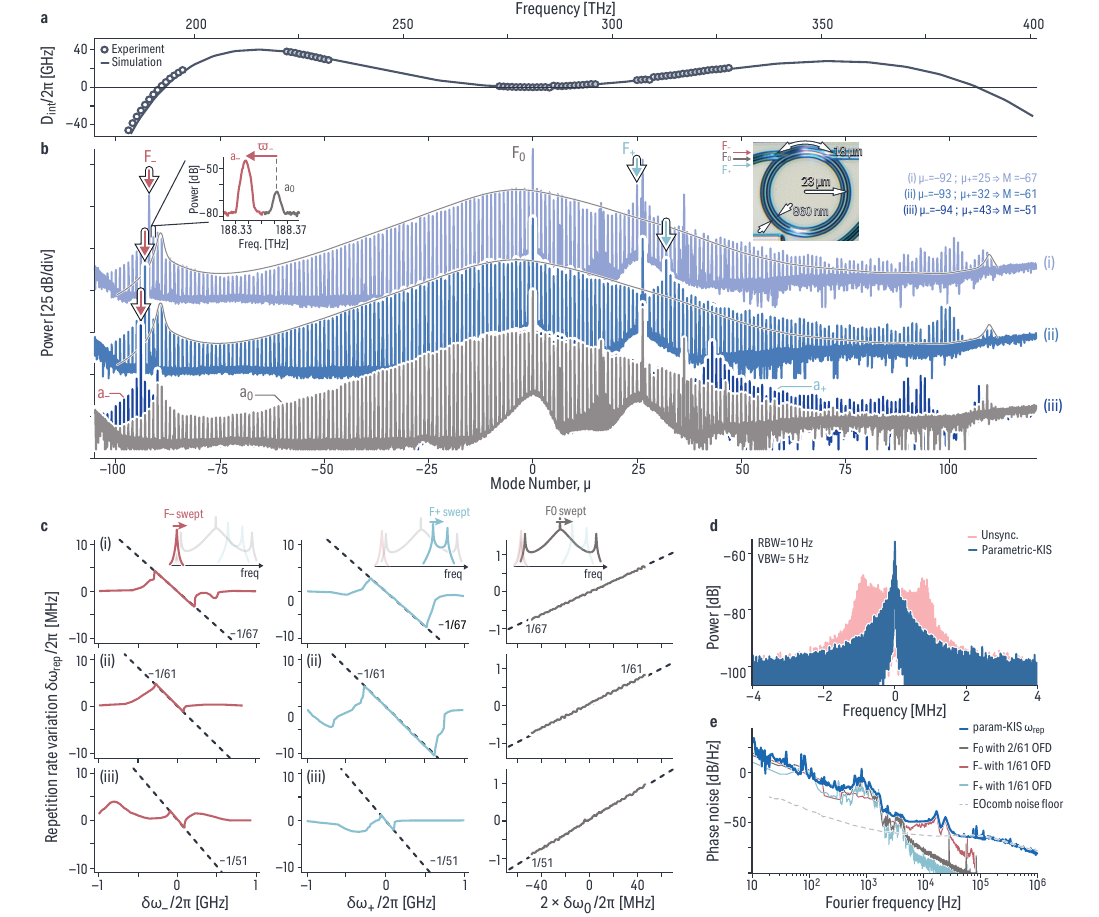}
        % aspect ratio 1.1986363636
        % word equivalent [300 / (0.5 * aspect ratio)] + 40
        % = 541
        \caption{\label{fig:3}
        \textbf{a} Integrated dispersion measurement (open circles) and simulation (solid line) for the microring resonator under study. %
        \textbf{b} Frequency comb optical spectra in parametric-KIS for different OFD factors, from top to bottom, $M=-67, -61$, and $-51$. 
        The left inset highlights the frequency offset $\varpi_-$ between the components $a_-$ and $a_0$. The soliton color $a_0$ is displayed in grey and remains the same for each $\mu_\pm$ configuration (grey envelope). The right inset shows a microscope image of the microring and its critical dimensions. %
        \textbf{c} Repetition rate disciplining for (i) $M=-67$,  -(ii) $-61$, and (iii) $-51$, where $\delta\omega_-$ (left), $\delta\omega_+$ (center), and $\delta\omega_0$ (right) is swept.  In the case of $\delta\omega_0$ sweep, we account for a factor 2 in the detuning due to the parametric nature of the process (see~\cref{eq:triple_pinning}). Only a small excursion within the parametric-KIS is shown for the main pump to avoid disrupting the DKS. %
        \textbf{d} Electrical spectra of the repetition rate beat when the DKS is free run ning (pink) and parametrically-synchronized (blue). RBW: resolution bandwidth; VBW: video bandwidth. The power is normalized to \qty{1}{~mW}, equivalent to dBm.
        \textbf{e} Phase noise of the repetition rate (blue thick) in the parametric-KIS regime, with the different noise contributions of each pump according to~\cref{eq:triple_pinning} displayed, highlighting the repetition rate optical frequency divided from the pump noises. The EOcomb apparatus (dash line) defines the noise floor of the repetition rate measurement. The power is normalized to the carrier's, equivalent to dBc/Hz.
        }
    \end{center}
\end{figure*}

\vspace{1em}
We now proceed to demonstrate parametric-KIS experimentally. We use an integrated microring resonator with a radius of \qty{R=23}{\um}, made of \qty{H=670}{\nm} thick \ce{Si3N4}, and a ring width of \qty{RW=860}{\nm} embedded in \ce{SiO2}. The bus waveguide is configured in a pulley fashion with a length \qty{L_c=18}{\um}, compensating for the coupling dispersion~\cite{MoilleOpt.Lett.2019} and allowing for efficient extraction of the entire comb. We pump the microring at a frequency \qty{\omega_0/2\pi\approx281.3}{\THz} with an on-chip power \qty{P_0=180}{mW} to generate an octave-spanning soliton microcomb at \qty{\omega_\mathrm{rep}/2\pi = 999.60376}{GHz} (\qty{\pm~10}{\kHz}), while using a cooler-laser at \qty{\approx308}{\THz} to thermally stabilize the system~\cite{ZhouLightSciAppl2019,ZhangOptica2019}. The comb exhibits two DWs [\cref{fig:3}b] at $\mu =-90$ (\qty{191.3}{\THz}) and $\mu = 108$  (\qty{389}{\THz}), due to the two zero crossings of $D_\mathrm{int}(\mu)$ [\cref{fig:3}a]. As described earlier, zero crossings lead to resonant modes $\mu_{\pm}$ that when pumped by auxiliary lasers, create colors $a_{\pm}$ with opposite phases $\varpi_{+} \approx -\varpi_{-}$. Such mode combinations can be found under dual pumping, where the $a_-$ color creates through nonlinear mixing a third color for which phase matching will be at $\mu_+$ (and vice-versa)~\cite{MoilleNat.Commun.2021,3wave}. In our experiment, such conditions, along with equipment compatibility, are met at $\{\mu_{-}; \mu_{+}\}= \{-92; 25\}$, $\{-93; 32\}$ and $\{-94; 43\}$, resulting in an OFD of $M=-67$, $-61$, and $-51$ [\cref{fig:3}b]. For all the experiments, the on-chip powers of the auxiliary pumps are set to \qty{P_-=1.25}{mW} and \qty{P_+=2.75}{mW}.  %
We record the DKS repetition rate using an electro-optic comb apparatus similar to refs~\cite{DrakePhys.Rev.X2019,MoilleNature2023,MoillearXiv2024} (see Supplementary Information S.5). The parametric-KIS is detected by recording the temporal trace of the EO-comb frequency down-converted $\omega_\mathrm{rep}$ while only one of the pump laser frequencies is swept. Once processed (see Supplementary Information S.5), this enables us to retrieve the dependence of $\omega_\mathrm{rep}$ on the laser detuning [\cref{fig:3}c].\\
\indent We confirm the parametric nature of such Kerr-induced synchronization since we obtain the $\omega_\mathrm{rep}$ entrainment, similar to the simulation, for $\delta \omega_\mathrm{rep}/\delta \omega_- = \delta \omega_\mathrm{rep}/\delta \omega_+ = 1/M$ %
 for each of the $\mu_\pm$ pairs under study. We note the difference in parametric KIS bandwidth from the $\mu_\pm$ auxiliary pumps, which we believe arises from the remaining $\kappa_\mathrm{ext}$ dispersion, impacting the $F_-$ and $F_+$ driving forces. 
 A unique feature of the parametric-KIS, deduced from \cref{eq:triple_pinning}, is the double contribution from the main pump. While in parametric-KIS,  we change the main pump frequency by $\delta\omega_0$---different from $\Delta\omega_0$ since the cooler pump thermally stabilizes the detuning from the resonance---and we observe a disciplining $\delta \omega_\mathrm{rep}/\delta \omega_0= -2/M$, as expected from the theory. \\
 \indent Finally, we characterize the parametrically synchronized DKS noise. Like direct-KIS, the pinned repetition rate shows significantly reduced noise compared to the untrapped case [\cref{fig:3}d]. Measurements of the repetition rate phase noise power spectral density (PSD) verify the parametric-KIS OFD. Using an optical frequency discriminator (see Supplementary Information S5), we find the repetition rate phase noise PSD matches the combined contribution of the three free-running lasers' phase noise PSD according to \cref{eq:triple_pinning} [\cref{fig:3}e], where the repetition rate detection is limited only by the EOcomb apparatus noise floor used for spectrally translating two DKS comb teeth into a detectable frequency. Since our OFD is competitive with state-of-the-art two-point locked microcombs for low-noise microwave generation~\cite{PappOpticaOPTICA2014c,SunNature2024a,KudelinNature2024}, we expect similar performances when the three lasers would be locked to frequency references.% which, after measuring the lasers phase noise with an optical frequency discriminator (see Supplementary Information S5), is in accordance with the three lasers' phase noise PSD contributions following~\cref{eq:triple_pinning} along with the aforementioned OFD factor~[\cref{fig:3}e].

%    ___                 _       
%   / __| ___  _ _   __ | | _  _ 
%  | (__ / _ \| ' \ / _|| || || |
%   \___|\___/|_||_|\__||_| \_,_|
                               
\vspace{0.5em}
\textit{Discussion---}%
In conclusion, we have demonstrated that harnessing the $\chi^{(3)}$ nonlinearity of a microresonator housing a dissipative Kerr soliton enables parametric Kerr-induced synchronization using two auxiliary lasers injected outside the soliton microcomb's frequency grid. As a result, the soliton is trapped in the field that is generated by the parametric interaction of different colors in the cavity. We have shown that this effect can be predicted by the multi-color formalism of the LLE, where the parametric interaction between the different waves gives rise to an additional force to the soliton color. We have numerically and experimentally demonstrated this effect, stabilizing the microcomb with auxiliary lasers outside the DKS microcomb's frequency grid. Additionally, other colors beyond the DKS can undergo parametric synchronization with suitable auxiliary pumping (see Supplementary Information S.7). Leveraging the group velocity binding for DKS stabilization through dual-pinning of another color~\cite{MoillearXiv2024}, this scheme offers enhanced flexibility for all-optical locking while relaxing the dispersion requirements for on-resonance operation compatible with pure quadratic $D_\mathrm{int}$. Our work presents the first prediction and demonstration of parametric synchronization of a DKS microcomb, opening new pathways for studying and applying the trapping of DKSs without direct actuation. It is important to note that parametric driving of solitons is not limited to $\chi^{(3)}$ systems~\cite{BruchNat.Photonics2021,EnglebertNat.Photonics2021}. Thus, the parametric-KIS of a DKS could be extended to other nonlinear orders, with potentially significant implications for the dual-pinning and noise of the frequency comb. Additionally, our work presents the potential for using multi-color solitons in metrology, harnessing their spectral extension~\cite{MoilleNat.Commun.2021} beyond the resonator's anomalous dispersion limit.
%     _        _   
%    /_\   __ | |__
%   / _ \ / _|| / /
%  /_/ \_\\__||_\_\

\vspace{1em}
\textit{Acknowledgment---}
G.M. K.S and M.E. acknowledge partial funding support from Marsden Fund of the Royal Society Te Apārangi [Grant No. 23-UOA-071]. G.M and K.S acknowledge partial funding support from the Space Vehicles Directorate of the Air Force Research Laboratory and the NIST-on-a-chip program of the National Institute of Standards and Technology. P.S and C.M. acknowledges support from the Air Force Office of Scientific Research (Grant No. FA9550-20-1-0357), the National Science Foundation (Grant No. ECCS-18-07272), and by a collaborative agreement 2022138-142232 and 2023200-142386 with the National Center for Manufacturing Sciences as a sub-award from US DoD cooperative agreement HQ0034-20-2-0007. 
We thank Logan Courtright and Usman Javid for insightful feedback. G.M. thanks T.B.M. 
%   ___  _  _     _      
%  | _ )(_)| |__ (_) ___ 
%  | _ \| || '_ \| |/ _ \
%  |___/|_||_.__/|_|\___/
                       
% \bibliography{Biblio}
%
    
%   _  _  _        _                             _           
%   ___               __  __        _   
%  / __| _  _  _ __  |  \/  | __ _ | |_ 
%  \__ \| || || '_ \ | |\/| |/ _` ||  _|
%  |___/ \_,_|| .__/ |_|  |_|\__,_| \__|
%             |_|                       
\clearpage
\onecolumngrid
\renewcommand{\appendixpagename}{\Large\centering Supplementary Information: \mytitle}
\appendix   
\appendixpage
\setcounter{figure}{0}
\renewcommand{\thesection}{S.\arabic{section}}
\renewcommand\thefigure{S.\arabic{figure}}    
\numberwithin{equation}{section}

\section{Derivation of the multi-color soliton coupled-LLE 
\label{supsec:multi_color_LLE}}
Here we present the derivation of the multi-color LLE to help the reader understand equation (3) of the manuscript. Such derivation can also be found in detail in ref~\cite{MoilleNat.Photon.2024} and particularly in ref~\cite{3wave}.

We start with the multi-pump LLE similar to ref~\cite{TaheriEur.Phys.J.D2017}:
\begin{equation}
\begin{split} 
    \label{sup_eq:mLLE}
    \frac{\partial a}{\partial t} &= 
        \left(-\frac{\kappa}{2} + i\Delta \omega_{\mathrm{0}}\right)a 
            + i\sum_\mu D_{\mathrm{int}}(\mu)A\mathrm{e}^{i\mu\theta} 
            - i\gamma  |a|^2 a + i\sqrt{\kappa_{\mathrm{ext}}P_{\mathrm{0}}} \\
        &+ i\sqrt{\kappa_\mathrm{ext}P_\mathrm{-}} \mathrm{e}^{i\left[\delta\omega_\mathrm{0} - \delta \omega_\mathrm{-} + D_\mathrm{int}(\mu_\mathrm{-})\right]t + i\mu_\mathrm{-}\theta} 
        + i\sqrt{\kappa_\mathrm{ext}P_\mathrm{+}} \mathrm{e}^{i\left[\delta\omega_\mathrm{0} - \delta \omega_\mathrm{+} +   D_\mathrm{int}(\mu_\mathrm{+})\right]t + i\mu_\mathrm{+}\theta}
    \end{split}
\end{equation}
with $\kappa$ and $\kappa_\mathrm{ext}$ the total and coupling loss rates, respectively, $\gamma$ the effective nonlinearity, $\theta$ and $\mu$ the azimuthal angle and component respectively, $A(\mu,t)$ the Fourier transform of $a(\theta,t)$ with respect to $\theta$, and  $D_\mathrm{int} = \omega_\mathrm{res}(\mu) - (\omega_0 + \omega_\mathrm{rep}\mu)$ the modified integrated dispersion with $\omega_\mathrm{rep}$ the repetition rate of the DKS, and $\mu$ the mode number relative to the main pumped mode. $P_0 = F_0^2/\kappa_\mathrm{ext}$, $\omega_0$, and $\Delta\omega_0$ are the power, frequency and detuning of the main pump creating the DKS, respectively; $\mu_\pm$ is the mode number of the positive (negative) auxiliary pump with power $P_\pm = F_\pm^2/\kappa_\mathrm{ext}$ and exhibiting an offset from the cloest soliton comb tooth $\varpi_\pm = \omega_\pm - \omega_0 - \mu\omega_\mathrm{rep}^{(0)}$. We defined the negative (positive) auxiliary pump such that $\mu_-<0$ ($\mu_+>0$).

We assume that three colors form from the three lasers injected in the cavity and therefore $a$ can be written as,
\begin{equation}
    \label{sup_eq:colors}
    a = a_{\mathrm{0}} + a_\mathrm{-} \mathrm{e}^{(i\varpi_\mathrm{-}t + i\mu_\mathrm{-}\theta)} + a_\mathrm{+} \mathrm{e}^{(i\varpi_\mathrm{+}t + i\mu_\mathrm{+}\theta)},
\end{equation}
Using \cref{sup_eq:colors} to substitute it in \cref{sup_eq:mLLE} and discarding any non phase-matched terms,
\begin{equation}
    \begin{split}
        &\frac{\partial a_\mathrm{0}}{\partial t} 
            + \frac{\partial a_\mathrm{-}}{\partial t}\mathrm{e}^{(i\varpi_\mathrm{-}t + i\mu_{\mathrm{aux}}\theta)} 
            + i\varpi_\mathrm{-}a_\mathrm{-}\mathrm{e}^{i\varpi_\mathrm{-}t + i\mu_{\mathrm{aux}}\theta}
            + \frac{\partial a_\mathrm{+}}{\partial t}\mathrm{e}^{(i\varpi_\mathrm{+}t + i\mu_{\mathrm{aux}}\theta)} 
            + i\varpi_\mathrm{+}a_\mathrm{+}\mathrm{e}^{i\varpi_\mathrm{+}t + i\mu_{\mathrm{aux}}\theta} = \\
        &\left(-\frac{\kappa}{2} + i\Delta \omega_\mathrm{0}\right)(a_\mathrm{0} 
                +  a_\mathrm{-}\mathrm{e}^{i\varpi_\mathrm{-}t + i\mu_\mathrm{-}\theta}
                +  a_\mathrm{+}\mathrm{e}^{i\varpi_\mathrm{+}t + i\mu_\mathrm{+}\theta}
            ) \\
        &+ i\sum_\mu D_\mathrm{int}(\mu)a_\mathrm{0}\mathrm{e}^{i\mu\theta}
            + i\sum_\mu D_\mathrm{int}(\mu)a_\mathrm{-}\mathrm{e}^{i\mu\theta}\mathrm{e}^{i\varpi_\mathrm{-}t} 
            + i\sum_\mu D_\mathrm{int}(\mu)a_\mathrm{+}\mathrm{e}^{i\mu\theta}\mathrm{e}^{i\varpi_\mathrm{+}t} \\
        &-i\gamma\Bigl[ \left(|a_\mathrm{0}|^2 + 2|a_\mathrm{-}|^2 + 2|a_\mathrm{+}|^2\right)a_\mathrm{0} 
        + \left(2|a_\mathrm{0}|^2 + |a_{\mathrm{-}}|^2 + 2|a_\mathrm{+}|^2\right)a_\mathrm{-} \mathrm{e}^{i\varpi_\mathrm{-}t + i\mu_\mathrm{-}\theta}
        + \left(2|a_\mathrm{0}|^2 + 2|a_{\mathrm{-}}|^2 + |a_\mathrm{+}|^2\right)a_\mathrm{+} \mathrm{e}^{i\varpi_\mathrm{+}t + i\mu_\mathrm{+}\theta} \Bigr]\\
        &-i\gamma \Bigl[
            2a_0^*a_{-}a_{+}e^{i(\mu_{-} + \mu_{+})\theta}e^{i(\varpi_- + \varpi_+)t} 
            + a_0^2a_+^*\mathrm{e}^{-i\varpi_\mathrm{+}t - i\mu_\mathrm{+}\theta} 
            + a_0^2a_-^*\mathrm{e}^{-i\varpi_\mathrm{-}t - i\mu_\mathrm{-}\theta} 
        \Bigr]\\
        &+ i\sqrt{\kappa_{\mathrm{ext}}P_\mathrm{0}}
        + i\sqrt{\kappa_{\mathrm{ext}}P_\mathrm{-}}\mathrm{e}^{i\varpi_\mathrm{-}t + i\mu_\mathrm{-}\theta}
        + i\sqrt{\kappa_{\mathrm{ext}}P_\mathrm{+}} \mathrm{e}^{i\varpi_\mathrm{+}t + i\mu_\mathrm{+}\theta}
    \end{split}
    \raisetag{6\normalbaselineskip}
\end{equation}
Separating the terms based on their phase $\varpi$ we get the system of three equations and noting that the system is set such that $\varpi_+ + \varpi_-$ is small before either $\varpi_-$ and $\varpi_+$,
% \begin{equation}
\begin{align}
    &\frac{\partial a_-}{\partial t} = 
        \left(- \frac{\kappa}{2} + i\varpi_-\right) a_- 
        + i\mathcal{D}_-(a_-)
        - i\gamma \left(2|a_0|^2 + |a_{-}|^2 + 2|a_{+}|^2\right)a_- 
        - i\gamma a_0^2a_{+}^*e^{-i[ M \theta + W t ]}
        + i\sqrt{\kappa_\mathrm{ext}P_-}\label{sup_eq:a-} \\
        &\frac{\partial a_0}{\partial t} = 
        \left(-\frac{\kappa}{2} + i\Delta\omega_0\right) a_0 
        + i\mathcal{D}_0(a_0) 
        - i\gamma \left(|a_0|^2 + 2|a_{-}|^2 + 2|a_{+}|^2\right)a_0 
        - 2i\gamma a_0^*a_{+}a_{-}e^{i[M\theta + W t]}
        + i\sqrt{\kappa_\mathrm{ext}P_0}\label{sup_eq:a0} \\
    &\frac{\partial a_+}{\partial t} = 
        \left(- \frac{\kappa}{2} + i\varpi_+\right) a_+ 
        + i\mathcal{D}_+(a_+)
        - i\gamma \left(2|a_0|^2 + 2|a_{-}|^2 + |a_{+}|^2\right)a_+ 
        - i\gamma a_0^2a_{-}^*e^{-i[M\theta + W t]}
        + i\sqrt{\kappa_\mathrm{ext}P_+}\label{sup_eq:a+}
\end{align}
% \end{equation}

\noindent with the dispersion operator terms $\mathcal{D}_0(a_0) = \sum_\mu D_\mathrm{int}(\mu)A_0 \mathrm{e}^{i\mu \theta}$; $\mathcal{D}_-(a_-) = \sum_\mu D_\mathrm{int}(\mu)A_-\mathrm{e}^{i(\mu - \mu_-) \theta}$; and $\mathcal{D}_-(a_-) = \sum_\mu D_\mathrm{int}(\mu)A_+\mathrm{e}^{i(\mu - \mu_+) \theta}$

The DKS $a_0$ therefore experiences another parametric driving force through the term $-2i\gamma a_0^*a_{+}a_{-}e^{i[W t + M\theta]}$, which has a phase offset from the DKS $W t = (\varpi_- + \varpi_+)t$ at an azimuthal mode $M=\mu_{-} + \mu_{+}$. Therefore, the system can becomes synchronized when $\varpi_- + \varpi_+$ is within the KIS bandwidth, with an OFD factor of $M$

The above system is the one solved to obtain the results presented in figure 2, which also explains the comb generation around the auxiliary pump at $\mu_\pm$. 

\section{Adler equation for the parametric trapping
\label{supsec:adler}}

Similar to the standard KIS, one could derive an Adler equation to understand the parametric trapping~\cite{MoilleNature2023}. 
From the definition of the parametric drive in ~\cref{sup_eq:a0}, its strongest azimuthal  mode will be at $M = \mu_{-} + \mu_{+}$ and will define the comb tooth of the DKS that will be captured. 
The variation of the phase offset of the $+$ and $-$ color relative to the DKS are defined by: 

\begin{align}
    \frac{\partial\varphi_+}{\partial t} = \varpi_+ = \omega_+ -  \left(\omega_0 + \mu_{+}\omega_\mathrm{rep}\right)\\
    \frac{\partial\varphi_-}{\partial t} = \varpi_- = \omega_- -  \left(\omega_0 + \mu_{-}\omega_\mathrm{rep}\right)
\end{align}
Therefore the phase offset of the parametric drive relative to the DKS is:
\begin{equation}
    \begin{split}
        \label{sup_eq:ceo_offset}
        \frac{\partial\Phi}{\partial t} = & \varpi_- + \varpi_+    \\\
                        =& -\left(\mu_{-} + \mu_{+}\right)\omega_\mathrm{rep} - 2\omega_0 +  \omega_- + \omega_+
    \end{split}
\end{equation}

From \cref{sup_eq:ceo_offset}, the OFD $M = \mu_{-} + \mu_{+}$ becomes obvious since in the synchronization regime $\partial\Phi/\partial t = 0$, leading to equation (4) of the manuscript:
\begin{equation}
    \label{sup_eq:triple_pinning}
    \omega_\mathrm{rep}^\mathrm{(pkis)} = \frac{\omega_{-} + \omega_{+} - 2\omega_0}{M}    \quad.
\end{equation}
From the above equation, we can derive the disciplining of the repetition rate against the frequency variation of either pump laser:

\begin{align}
    \frac{\partial\omega_\mathrm{rep}}{\partial\omega_-} &= +\frac{1}{\mu_+ + \mu_-}\label{sup_eq:ofd_-}\\
    \frac{\partial\omega_\mathrm{rep}}{\partial\omega_0} &= -\frac{2}{\mu_+ + \mu_-}\label{sup_eq:ofd_0}\\
    \frac{\partial\omega_\mathrm{rep}}{\partial\omega_+} &= +\frac{1}{\mu_+ + \mu_-}\label{sup_eq:ofd_+}
\end{align}
which correspond to the experimental results shown in Figure 3c where $\mu_- + \mu_+<0$. The factor of 2 in the OFD relative to the main pump $\omega_0$  in~\cref{sup_eq:ofd_0} and the same OFD sign for $\omega_\pm$ in~\cref{sup_eq:ofd_-,sup_eq:ofd_+} regardless of the sign of $\mu_\pm$ are self-consistent and are clear signatures of the parametric nature of such Kerr-induced synchronization.\\

\cref{sup_eq:ceo_offset} links the repetition rate variation in time with the phase offset, assuming other terms are independent of time (detuning, optical frequencies): 
\begin{equation}
    \label{sup_eq:d2phidt2}
    \frac{\partial^2\Phi}{\partial t^2} = -\left(\mu_{-} + \mu_{+}\right) \frac{\partial\omega_\mathrm{rep}}{\partial t}
\end{equation}
\vspace{1em}

In the approximation of small amplitude auxiliary driving force $F_\pm$, their introduction has no impact on the soliton repetition rate~\cite{MoilleNature2023}, which then can be written as: 

\begin{equation}
    \frac{\partial\omega_\mathrm{rep}}{\partial t } = -\frac{1}{E_\mathrm{dks}} \sum_\mu D_1(\mu) \left(\frac{\partial A_0(\mu)}{\partial t}A_0^*(\mu) + A_0(\mu)\frac{\partial A_0^*(\mu)}{\partial t}  \right)
\end{equation}

Using the modal expansion of the Lugiato-Lefever equation and achromatic dispersive coupling and losses, and that only the terms at $M = \mu_{-} + \mu_{+}$ and $\mu = 0$ will be significant and that the auxiliary pumps are not depleted:
\begin{equation}
    \label{sup_eq:dwrepdt}
    -\frac{1}{\kappa}\frac{\partial\omega_\mathrm{rep}}{\partial t } = \frac{2D_1}{E_\mathrm{dks}} \sqrt{E_0(M)\frac{4\gamma^2 P_0P_+P_- \kappa_\mathrm{ext}^{3}}{\kappa^2}  }\sin(\Phi) - \omega_\mathrm{rep} - D_1 \left( K_0 + K_\mathrm{NL} \right)
\end{equation}
with $E_0(M) = |A_0(M)|^2$, where the approximation $D_1(\mu=0)\approx D_1(\mu = M) = D_1$ since the resonator is weakly dispersive. The pump frequency shift and the self-phase modulation frequency shift are: 
\begin{align}
    K_0 &= 2\sqrt{\kappa_\mathrm{ext}}\frac{P_0}{\kappa E_\mathrm{dks}}\\
    K_\mathrm{NL} &= \frac{\gamma}{\kappa E_\mathrm{dks} D_1 }\sum_\mu D_1(\mu)\left(A^*\left( \mu \right) \sum_{\alpha,\beta}A(\alpha)A^*(\beta)A(\alpha-\beta+\mu)  - c.c \right)
\end{align}

\noindent which from the steady state form of~\cref{sup_eq:dwrepdt} leads to $D_1 \left( K_0 + K_\mathrm{NL} \right)\approx \omega_\mathrm{rep}$ since $E_\mathrm{kis}\ll 1$ .

Using \cref{sup_eq:ceo_offset,sup_eq:d2phidt2} we obtain the Adler equation: 

\begin{equation}
    \label{sup_eq:parametric_adler}
    \beta\frac{\partial\Phi^2}{\partial\tau^2} + \frac{\partial\Phi}{\partial\tau} = \Delta  - \sin(\Phi)
\end{equation}

\noindent with the normalized time $\tau = \Omega_\mathrm{pkis} t = 2 M D_1 E_\mathrm{pkis}t$ with $\Omega_\mathrm{pkis}$ the synchronization bandwidth and the normalized KIS energy $E_\mathrm{pkis} = \sqrt{4\gamma^2 E_0(M)P_0P_+P_-\kappa_\mathrm{ext}^{3}/{\kappa^2}}/E_\mathrm{dks}$, the damping $\beta = \Omega_{pkis}/\kappa$, and the detuning $\Omega_\mathrm{pkis}\Delta =  \left( \omega_- + \omega_+ -2\omega_0 \right) =  W$. Hence, if $\Delta<1$, the DKS becomes synchronizes and acquires a phase slip $\partial\varphi_\mathrm{dks}/\partial t$ which compenstae for $W$ such that $\partial\Phi/\partial\tau$ remains null. Since the main pump pins one comb tooth, this phase slip transduces into a group velocity shift changing the DKS repetition rate.

\section{Summary of parameters
\label{supsec:parameters}}

\begin{figure}[h]
    \centering
    \includegraphics[width = \textwidth]{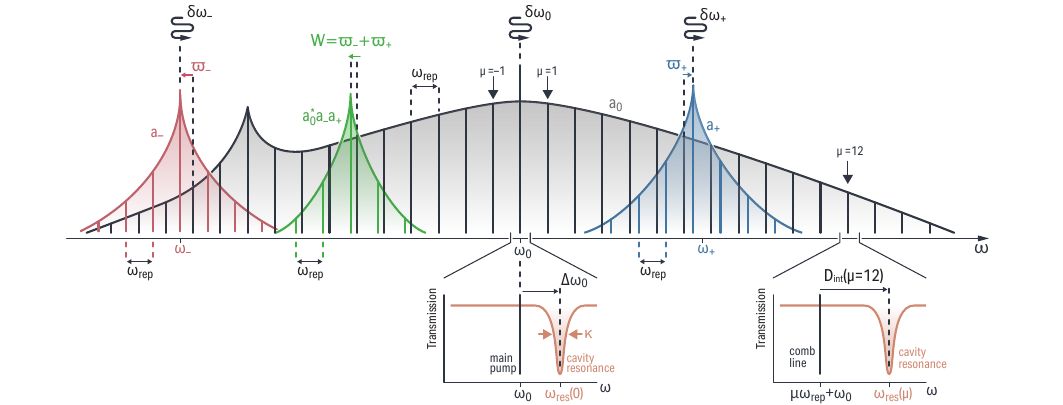}
    \caption{\label{sup_fig:params} Graphical representation of the frequency parameters, $\omega_0$, $\omega_\pm$, $\mu$, $D_\mathrm{int}(\mu)$, $\omega_\mathrm{rep}$, $\omega_\mathrm{res}(\mu)$, $\kappa$, $\varpi_\pm$, and $W$.
    }
    \vspace{2em}
\end{figure}

The different frequency parameters used in the manuscript are summarized in ~\cref{sup_fig:params}. The main pump $\omega_0$ is detuned from its cold cavity resonance $\omega_\mathrm{res}(0)$ by $\Delta\omega_0 = \omega_\mathrm{res}(0) - \omega_0$. The resonances have a linewidth $\kappa$, which accounts for both intrinsic and extrinsic (coupling) losses $\kappa_\mathrm{ext}$. The frequency variation or tuning of the main pump is $\delta\omega_0$, which in the LLE can be related to $\Delta\omega_0$. However, since experimentally we use a cooler pump to thermally stabilize the resonator, $\Delta\omega_0$ does not change linearly with $\delta\omega_0$. The auxiliary pumps are located around comb teeth $\mu_\pm$ away from the main pump, such that $\mu_-<0$ and $\mu_+>0$. They are frequency offset from their closest comb teeth by $\varpi_\pm$. The parametric interaction between the three colors at play $a_-$, $a_+$, and $a_0$ pumped by the auxiliary pump and main pump respectively generates an idler centered at the frequency $\omega_0 + M \omega_\mathrm{rep} + W$, with $M = \mu_- + \mu_+$ and $W=\varpi_- + \varpi_+$. In other words, the idler occurs around the comb tooth at mode $M$ and is detuned from it by $W$. When $W$ is small enough to be within the KIS bandwidth, parametric synchronization occurs where the repetition rate of the DKS is defined by $\omega_\mathrm{rep} = M^{-1}\left( \omega_- + \omega_+ - 2\omega_0 \right)$.

\section{Microring resonator design
\label{supsec:microring_design}}

The microring resonator's chip consists of a stack of silicon, silicon dioxide (\ce{SiO2}), silicon nitride (\ce{Si3N4}), and \ce{SiO2}. The photonic fabrication is performed in a commercially available foundry, similar to the chips presented in~\cite{MoilleNature2023}. The photonics layer consists of \qty{H=670}{\nm} thick \ce{Si3N4}, embedded in \ce{SiO2}, enabling light guiding. The ring resonator has a width of \qty{RW=860}{\nm} and a radius of \qty{RR=23}{\um}. The waveguide that couples to the ring has a width of \qty{W=460}{\nm} and is separated from the ring by a gap of \qty{G=500}{\nm} across a \qty{L_c=17}{\um} length in a pulley fashion~\cite{MoilleOpt.Lett.2019}. The light is injected and collected through the facet edge. Light injection is efficient, with low insertion losses of about \qty{3}{\dB} per facet, thanks to the inverse tapering of the waveguide down to a width of \qty{250}{\nm}.

% \clearpage
\section{Experimental setup
\label{supsec:exp_setup}}

\begin{figure}[h]
    \centering
    \includegraphics[width = \textwidth]{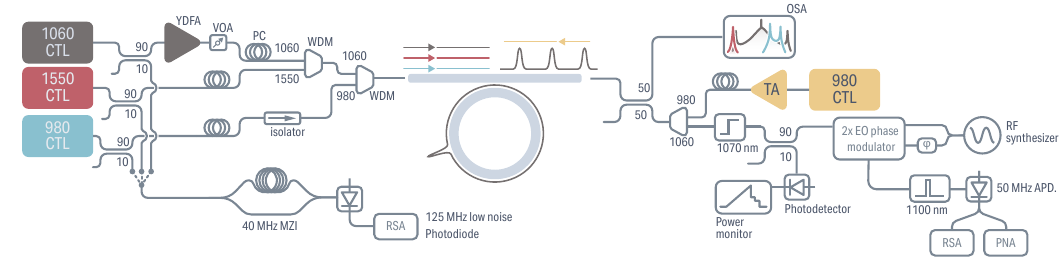}
    \caption{\label{sup_fig:setup}
    Experimental setup. The \qty{1060}{\nm} continuously tunable laser (CTL), amplified with a ytterbium doped-fiber amplifier (YDFA) and power-adjusted with a variable optical attenuator (VOA),  generates the DKS thanks to the thermal stabilization of the microring resonator by the \qty{980}{\nm} CTL cooler pump laser. The main pump laser along with the \qty{1550}{\nm} and \qty{980}{\nm} auxiliary CTL lasers are combined by a wavelength demultiplexer (WDM). An isolator is present in the \qty{980}{\nm} path to avoid the cooler feeding into the auxiliary pump  CTL. The soliton step is detected by filtering the main pump from the comb power detection. The comb spectrum is detected by an optical spectrum analyser (OSA). The repetition rate is detected thanks to the creation of an electro-optic comb (EOcomb) of two adjacent DKS comb teeth, which is then filtered at the overlapping frequency of the two spectrally translated DKS comb teeth, and detected with an avalanche photodiode (APD). The signal is then sent to a real-time spectrum analyser (RSA) and phase noise analyzer (PNA) for analysis. The laser noise is measured thanks to a \qty{40}{\MHz} Mach-Zehnder interferometer (MZI) acting as an optical frequency discriminator and detected with a \qty{125}{\MHz} low noise photodiode and a RSA. 
    \vspace{1em}
    }
\end{figure}

The setup [\cref{sup_fig:highercolor}] consists of the previously described microring resonator, fiber elements, detector, analysis instruments, and four lasers. The \qty{1060}{\nm} continuously tunable laser (CTL) is amplified through a ytterbium-doped fiber amplifier (YDFA) to achieve the required \qty{>150}{\mW} of in-fiber power to generate a frequency comb. A polarization controller (PC) sets the input light into transverse-electric (TE) polarization to pump the fundamental TE mode of the microring resonator. A variable optical attenuator (VOA) controls the power to reach the single soliton state. A \qty{980}{\nm} CTL, amplified through a tapered amplifier (TA) up to \qty{400}{\mW}, is used in a cross-polarized, counter-propagative manner to thermally stabilize the microring while minimizing its nonlinear interaction with the DKS. This thermal stabilization, similar to opto-mechanical cooling—hence the term \textit{cooler} pump laser—enables adiabatic frequency tuning to the single soliton state, detected by measuring the comb power while notching the main pump, while also recording the spectrum with an optical spectrum analyzer (OSA).

The repetition rate $\omega_\mathrm{rep}$ of the microcomb is detected using two cascaded electro-optic (EO) phase modulators driven by an amplified microwave synthesizer. By appropriately setting the driving frequency, the EO comb of two adjacent DKS comb teeth overlaps within the bandwidth of the avalanche photodiode (APD) used for detection. The repetition rate is determined as $\omega_\mathrm{rep} = N_\mathrm{eo}\omega_\mathrm{eo} \pm \omega_\mathrm{beat}$, where $N_\mathrm{eo} = 56$ is the number of EO comb teeth, $\omega_\mathrm{eo}$ is the EO-comb microwave frequency, and $\omega_\mathrm{beat}$ is the beat note frequency. The $\pm$ sign arises from the overlap (or not) of the EO-combed OFC comb teeth. We record the signal through the real-time spectrum analyzer (RSA), extracting either the spectrum, or the in-phase and quadrature (IQ) signal of $\omega_\mathrm{beat}$. The IQ signals enable, while sweeping one of the pump lasers, the reconstruction of a spectrogram from which the instantaneous frequency of the beat note can be extracted. Calibration of the laser detuning is obtained through a heterodyne beat against a local oscillator, in this case, an additional laser close in wavelength.

For noise analysis, we use a phase-noise analyzer to characterize $\langle\omega_\mathrm{beat}\rangle$. Since it carries both the DKS comb noise $\langle\omega_\mathrm{rep}\rangle$ and the multiplied EO-comb microwave synthesizer noise $N\langle\omega_\mathrm{eo}\rangle$, our noise floor analysis is limited by the EO-comb.

Finally, we measure the frequency noise of every CTL at play in this experiment using an optical frequency discriminator based on a \qty{\omega_{MZI} = 2\pi \times 40}{\MHz} fiber Mach-Zehnder interferometer (MZI) and a \qty{125}{\MHz} low-noise photodiode. The signal is then recorded with the RSA. From the electrical spectrum, we obtain the frequency spectral density of noise of each laser $S_\omega = \omega_{MZI}  \sin^{-1}\left(\frac{S_\mathrm{rsa}}{A\pi} \right)^ 2$ with $A$ the maximum amplitude of the interference pattern, and $S_\mathrm{rsa}$ the RSA signal. Converting the frequency noise PSDs to phase noise PSDs allows for the verification of the repetition rate triple-pinning. We compare the weighted sum of the laser phase noise PSDs, using equation (4), against direct measurements of the repetition rate phase noise (see Figure 3f of the main manuscript).

\clearpage
\section{LLE simulations for different parametric-KIS OFDs}
\label{supsec:trapp_differentmu}

\begin{figure}[h]
    \centering
    \includegraphics[width = \textwidth]{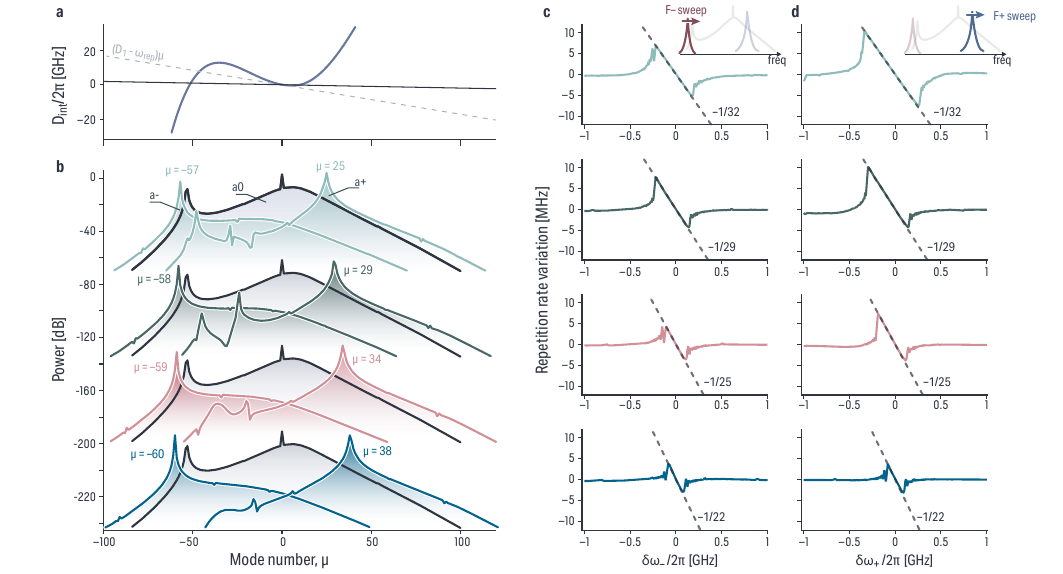}
    \caption{\label{sup_fig:trapp_differentmu}
    \textbf{a} Modified integrated dispersion, similar to the one displayed in Fig.~2. 
    \textbf{b} Optical frequency comb spectra of the DKS (grey center) and the auxiliary pumped color for different combination of $\mu_\pm$ yielding different OFD factor $M=\mu_- + \mu+=$.
    \textbf{c} Repetition rate disciplining from the parametric-KIS on the DKS $a_0$, whose slope is following the OFD factor $M=-32$, $-29$, $-25$, and $-22$ and is consistent with the theory presented.% 
    }
    \vspace{2em}
\end{figure}

We extend the LLE simulations presented in Fig.~2 to different sets of auxiliary pump modes $\mu_\pm$ [\cref{sup_fig:trapp_differentmu}]. Using combinations of modes ${\mu_{-}; \mu_{+}}= {-57; 25}$, ${-58; 29}$, ${-59; 34}$, and ${-50; 38}$, corresponding to the parametric-KIS at mode $M=-32$, $-29$, $-25$, and $-22$ respectively. Similar to the study presented in the main text, we vary the frequency of the auxiliary pump $\delta\omega_\pm$ and observe the repetition rate disciplining of the DKS, with slopes consistent with the OFD factor M [\cref{sup_fig:trapp_differentmu}c]. Importantly, the same microcomb (\textit{i.e} given pump frequency and resonator dispersion) can be efficiently on-resonance for parametric synchronization at multiple OFD values. This contrasts with direct-KIS, where efficient on-resonance synchronization occurs only at the dispersive wave (not limited for off-resonance operation). Such flexibility highlights the advantage of the parametric nature of this synchronization scheme.

% \clearpage
\section{Trapping of other colors
\label{supsec:trapp_highercolor}}

The system of equations derived in \cref{sup_eq:a0,sup_eq:a-,sup_eq:a+} can be easily expanded to other colors by renormalizing it to the $+1$ color instead of the DKS:
\begin{equation}
    a = a_0 + a_{+1}\mathrm{e}^{i\left[\mu_{+1}\theta + \varpi_{+1}t\right]} +a_{+2} \mathrm{e}^{i\left[\mu_{+2}\theta + \varpi_{+2}t\right]}
\end{equation}

\noindent from expansion of the multi-pumped LLE, yields:
\small\begin{align}
    &\frac{\partial a_0}{\partial t} = 
        \left( -\frac{\kappa}{2} + i\Delta\omega_0\right) a_0 
        + i\mathcal{D}_0(a_0) 
        - i\gamma \left(|a_0|^2 + 2|a_{+1}|^2 + 2|a_{+2}|^2\right)a_0 
        - i\gamma a_{+1}^2a_{+2}^* e^{-i[M\theta + W t]}
        + i\sqrt{\kappa_\mathrm{ext}P_0}\label{sup_eq:high_a0} \\
    &\frac{\partial a_{+1}}{\partial t} = 
        \left( -\frac{\kappa}{2} + i\varpi_{+1}\right) a_{+1} 
        + i\mathcal{D}_{+1}(a_{+1})
        - i\gamma \left(2|a_0|^2 + |a_{+1}|^2 + 2|a_{+2}|^2\right)a_{+1}
        - 2i\gamma a_{+1}^*a_{0}a_{+2}e^{i[ M \theta + W t ]}
        + i\sqrt{\kappa_\mathrm{ext}P_{+1}}\label{sup_eq:high_a+1} \\
    &\frac{\partial a_{+2}}{\partial t} = 
        \left(- \frac{\kappa}{2} + i\varpi_{+2}\right) a_{+2} 
        + i\mathcal{D}_{+2}(a_{+2})
        - i\gamma \left(2|a_0|^2 + 2|a_{+1}|^2 + |a_{+2}|^2\right)a_{+2} 
        - i\gamma a_{+1}^2a_{0}^* e^{-i[M\theta + W t]}a
        + i\sqrt{\kappa_\mathrm{ext}P_{+2}}\label{sup_eq:high_a+2}
\end{align}
\normalsize
with $M = \mu_{+2} -2\mu_{+1}$ and $W = \varpi_{+2} - 2\varpi_{+1}$.
\vspace{1em}

% \begin{wrapfigure}{l}{0.5\textwidth}
\begin{minipage}[c]{0.5\textwidth}
    \begin{center}
        \includegraphics{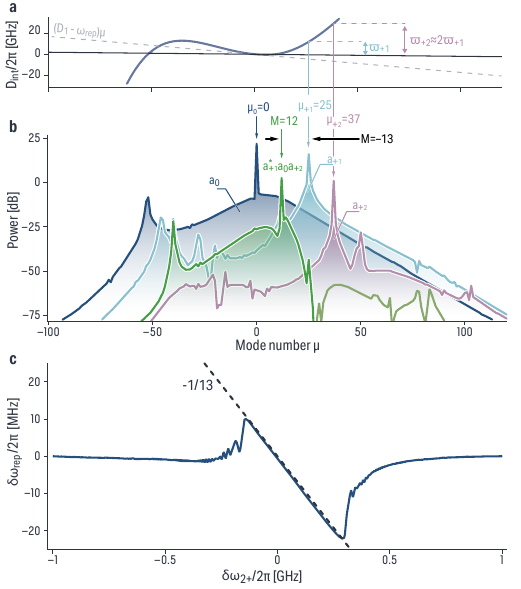}
    \end{center}
\end{minipage}
\begin{minipage}{0.5\textwidth}
    \captionof{figure}{\label{sup_fig:highercolor} \textbf{Parametric trapping of other colors instead of the soliton.} 
    \textbf{a} Integrated dispersion under consideration, similar to the one used in Fig. 2. %
    \textbf{b} Optical frequency comb spectra of the colors of interest, with $a_0$ (dark blue) the DKS and $a_{+1}$ (light blue) and $a_{+2}$ (purple) the auxiliary pumped colors. The parametric trapping term is shown in green. \textbf{c} Repetition rate disciplining from the parametric-KIS on the color $a_{+1}$, whose OFD factor $M=-13=\mu_{+2}-2\mu_{+1}$ is consistent with the theory presented in \cref{supsec:trapp_highercolor}.%
    }
\end{minipage}

% \end{wrapfigure}

Similar to~\cref{supsec:adler}, one can use~\cref{sup_eq:high_a+1} to reach the same normalized Adler equation: 
\begin{equation}
    \beta\frac{\partial\Phi^2}{\partial\tau^2} + \frac{\partial\Phi}{\partial\tau} = \Delta_{+1} +  \sin(\Phi)
\end{equation}

\noindent this time with $\tau=$ $E_\mathrm{pkis} = $, $\Delta_{+1} = W/\Omega_\mathrm{pkis}$.  Hence, the OFD factor now becomes $M = \mu_{+2} - 2\mu_{+1}$. This system can also be understood in the same fashion as presented in \cref{supsec:multi_color_LLE} if the phase and azimuthal offset of each color are normalized to the $+1$ color instead of the DKS.

We simulate this system, with results shown in~\cref{sup_fig:highercolor} that are consistent with the theory presented above.

\section{Direct-KIS vs Parametric-KIS}

\begin{figure}[H]
    \centering
    \includegraphics{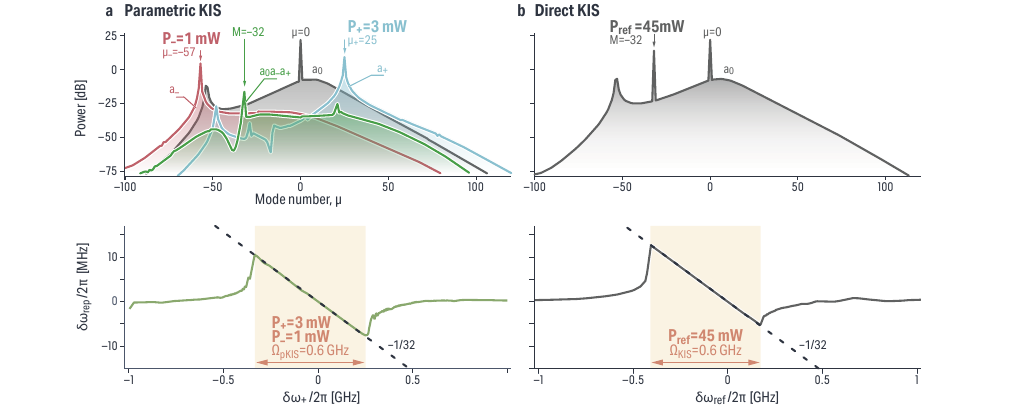}
    \caption{\label{sup_fig:dKIS_vs_pKIS}
    Comparison of parametric KIS and direct KIS \textbf{a} System from Fig.~2 in the main text. Parametric-KIS operation with \qty{P_-=1}{\mW} and \qty{P_+=3}{\mW} creates an idler at $M=-32$, disciplining the DKS repetition rate within a \qty{\Omega_\mathrm{pkis}\approx 0.6}{\GHz} window. \textbf{b} Same system as in a, using direct-KIS with a reference laser power for direct comparison with the parametric-KIS case. Off-resonance operation yields lower intracavity field power than parametric-KIS, resulting in a much larger reference power needed \qty{P_\mathrm{ref}=45}{\mW}  to obtain a direct KIS window \qty{\Omega_\mathrm{kis}\approx 0.6}{\GHz}, similar to the parametric-KIS case. 
    }
\end{figure}

We compare direct-KIS~\cite{MoilleNature2023} and parametric-KIS efficiency under identical conditions [\cref{sup_fig:dKIS_vs_pKIS}] using the same dispersion and microcomb from Figure 2 in the mLLE. For parametric-KIS, reference pumps with powers \qty{P_- =1}{\mW}, \qty{P_+=3}{\mW} at $\mu_-=57$ and $\mu=+25$ respectively create a parametric tone at $M=-32$, yielding a \qty{\Omega_\mathrm{pkis}\approx0.6}{\GHz} synchronization bandwidth. For direct-KIS at off-resonant $M=-32$, achieving similar bandwidth \qty{\Omega_\mathrm{kis}\approx0.6}{\GHz} requires \qty{P_\mathrm{ref} = 45}{\mW}---significantly higher than parametric-KIS auxiliary powers.
From \cref{sup_eq:parametric_adler}, parametric-KIS energy scales with $\sqrt{4\gamma^2 P_0 P_-P_+\kappa_\mathrm{ext}^3}$, while direct-KIS scales with $\sqrt{P_\mathrm{ref}\kappa_\mathrm{ext}}$ at fixed comb tooth power $E_0(M)$, mode number $M$, DKS energy $E_\mathrm{dks}$, and loss rate $\kappa$. On-resonant operation enables parametric-KIS to optimize synchronization energy more efficiently than direct-KIS.

% \bibliography{Biblio}
\end{document}